\RequirePackage{fix-cm}
\documentclass[smallextended]{svjour3}
\smartqed
\usepackage{amsmath,amssymb,paralist,subfigure,graphicx,cite}
\usepackage{algorithm2e}
\usepackage{color}

\def\mb{\mathbf}

\def\mc{\mathcal}

\begin{document}
	
	\title{How Clustering Affects the Convergence of Decentralized Optimization over Networks: A Monte-Carlo-based Approach
	}
	\author{Mohammadreza Doostmohammadian \and Shahaboddin Kharazmi \and
		Hamid R. Rabiee
	}
	
	\institute{ M.  Doostmohammadian \at
		Department of Mechatronics, Faculty of Mechanical Engineering, Semnan University, Semnan, Iran. \\
		Tel.: +98-23-31533429\\
		\email{doost@semnan.ac.ir}		
		\and		
		 S. Kharazmi \at
		 Faculty of Mechanical Engineering, Semnan University, Semnan, Iran. \\
		 Tel.: +98-23-31533429\\
		 \email{kharazmi@semnan.ac.ir}
		\and		
		H. R. Rabiee \at
		Department of Computer Engineering, Sharif University of Technology, Tehran, Iran. \\ 
		\email{rabiee@sharif.edu} 
		\and
		The first two authors  M.  Doostmohammadian and S. Kharazmi contributed equally to this paper.
		}
	
	\date{Received: date / Accepted: date}

	\maketitle
	
	\begin{abstract}
	Decentralized algorithms have gained substantial interest owing to advancements in cloud computing, Internet of Things (IoT), intelligent transportation networks, and parallel processing over sensor networks. The convergence of such algorithms is directly related to specific properties of the underlying network topology. Specifically, the clustering coefficient is known to affect, for example, the controllability/observability and the epidemic growth over networks.
	In this work, we study the effects of the clustering coefficient on the convergence rate of networked optimization approaches.   In this regard, we model the structure of large-scale distributed systems by random scale-free (SF) and clustered scale-free (CSF) networks and compare the convergence rate by tuning the network clustering coefficient. This is done by keeping other relevant network properties (such as power-law degree distribution, number of links, and average degree) unchanged. Monte-Carlo-based simulations are used to compare the convergence rate over many trials of SF graph topologies. Furthermore, to study the convergence rate over real case studies, we compare the clustering coefficient of some real-world networks with the eigenspectrum of the underlying network (as a measure of convergence rate). The results interestingly show higher convergence rate over low-clustered networks. This is significant as one can improve the learning rate of many existing decentralized machine-learning scenarios by tuning the network clustering.
		
	\end{abstract}
\section{Introduction}	
	In the realm of modern cloud-based and distributed systems \cite{bowman2019consensus}, the need for faster machine learning (ML) and artificial intelligence (AI) solutions over networked systems and complex social networks is essential. This further finds application in distributed transportation networks, simulation-based automotive network optimization \cite{shamsderakhshan2014turbocharger}, platooning of networked vehicles \cite{abolfazli2023minimum}, and vehicular Ad-Hoc (communication) networks \cite{wang2016complex}.
Network characteristics (such as clustering coefficient, degree distribution, network density, and average shortest path length) play crucial roles in understanding and influencing control, optimization, and learning processes over networks. Many works in the literature are devoted to this research area. It is claimed in \cite{karrer2011stochastic} that heterogeneous degree distributions can influence the efficiency of optimization algorithms. As reported by \cite{kandhway2016optimal}, identifying high-degree nodes may lead to better resource allocation and faster convergence over networks. Another interesting work shows that higher-density networks can facilitate faster information dissemination and control responses \cite{liu2016impacts}. However, excessively dense networks may lead to increased computational costs for control strategies. Similarly, dense networks can offer diverse paths for optimization algorithms \cite{elbayoumi2022edge}, while the challenge lies in balancing the benefits of increased connectivity with the potential for information overload or redundant pathways. On the same research line, dense networks are known to provide richer datasets for learning algorithms \cite{gures2022machine}. However, balancing density with diversity is crucial to prevent overfitting. Graph signal processing \cite{leus2023graph} is another interesting research direction that studies the system structures modelled by graphs capturing their complex interactions and serving as the basis for a theory of processing signals.
Among the network properties, \textit{clustering} is the main focus of this paper and is known to have a crucial role in the control and diffusion processes over networks \cite{abbasi2007survey}. It is shown that one can manage the network controllability/observability by tuning its clustering \cite{lcss_cluster,JCN}. Similarly, control of epidemics  \cite{coupechoux2014clustering,snam2} and spread of contagious disease over networks \cite{miller2009spread,snam} are shown to be directly related to the network clustering.

The role of network structure is also discussed in distributed optimization and machine learning literature; for example, see \cite{nedic2018network,yang2010distributed,verbraeken2020survey}. The works \cite{rossi2020distributed,pirani2023graph} study network diameter\footnote{The diameter represents the maximum shortest path between any pair of nodes.} and claim that smaller diameter generally facilitates faster communication and convergence, influencing the scalability of distributed optimization and consensus algorithms. Also, highly connected networks are associated with faster convergence, while low connectivity may result in slower information diffusion and potentially hinder the optimization process \cite{dsvm}. This is denoted by \textit{algebraic connectivity} and is known to directly affect the consensus rate of convergence \cite{SensNets:Olfati04,dsvm}.   
Some other works \cite{xin2021fast,assran2019stochastic} show that exponential networks with specific hop-based linking lead to fast linear convergence as compared to their randomly-connected counterparts. The convergence rate of the existing decentralized optimization and learning methods can also be tuned by momentum-based solutions  \cite{xin2019distributed,barazandeh2021decentralized,jin2022momentum} by adding momentum term (also referred to as the \textit{heavy ball}) to the nodes' information updates. The use of surrogate functions to enhance the convergence rate (as a function of network connectivity) is proposed by \cite{sun2022distributed}. In \cite{nazari2021adaptive} the bound on the convergence is shown to follow path-length of the comparator sequence and the network connectivity. Designing \textit{robust} graph topologies for fault-tolerant distributed optimization is proposed by \cite{wang2023resilient}. Some other works \cite{ogiwara2015maximizing,alenazi2014cost} are devoted to increasing the algebraic connectivity for faster synchronization and consensus over networks.  Among different network properties, there is a gap in the literature on how clustering affects the optimization convergence rate.

In this paper, we consider random SF and CSF networks and tune their clustering coefficient by managing the \textit{triangle formations} within the network. We adopt the distributed learning algorithm proposed in \cite{dsvm,ddsvm} to study the convergence rate versus the network clustering coefficient. First, a measure of the convergence rate of the algorithm as a function of system eigenvalue is given. Next, this is used for checking the rate of convergence versus different network characteristics. In general, there is no analytical solution to prove the relation between the clustering coefficient and network eigenspectrum (i.e., the problem is generally NP-hard to solve analytically); thus, we perform Monte-Carlo simulations and run the optimization algorithm over many trials of networks to compare its convergence versus clustering. Other key network parameters including the average node degree, network linking, and power-law degree distribution are set unchanged to solely study the effect of the clustering coefficient. Other than synthetic networks, this is also studied over real-world network cases and the clustering coefficient is compared with the convergence rate measure for different networks.
Our results interestingly show a meaningful relation between the clustering and the rate of convergence. Specifically, by increasing the clustering coefficient of CSF networks the convergence rate of the optimization/learning algorithm decreases. This is practically significant as one can improve the convergence rate by tuning the clustering coefficient of large-scale networks. 

The findings in this paper can be applied to solve real-world problems over complex networks. One example is the fast outbreak sensing problem which aims to find an optimal observer set such that the outbreaks over complex networks can be timely detected by monitoring the state of the nodes in the observer set \cite{liu2024fast}. Another example is the containment of diffusion in a network immunization perspective \cite{10336828}. In this case, the problem is to suppress the giant connected component of a complex network by removing as fewer nodes as possible, so that the intervention of the transmission could be achieved by only a few resources. Similarly, the diffusion-source-inference problem can be considered as another application, where the idea is to find the real source of an undergoing or finished diffusion \cite{liu2023diffusion}. 

The rest of the paper is organized as follows. Section~\ref{sec_graph} sets the graph theoretic background.  Section~\ref{sec_conv} and~\ref{sec_case} present the main results on the convergence rate versus clustering over synthetic and real networks. Section~\ref{sec_conc} concludes the paper.

Notations: Here, we summarize some main notations used in the paper.
	\begin{table}[hbpt!]
		\centering
		\caption{Table of notations.}
		\begin{tabular}{|c|c|}
			\hline
           $\mc{G}$& graph topology of the network \\
           \hline $\lambda_2(\cdot)$ &  second largest eigenvalue of matrix \\
           \hline
			$\mc{C}$& global clustering coefficient \\
			\hline $d$ &  average node degree \\
			\hline
			$\mc{C}_i$ & local clustering at node $i$ \\
			\hline $d_i$ &  degree of node $i$ \\
			\hline
			$n$ & size of the network (number of nodes) \\
			\hline $\mc{T}_i$ &  number of triads that contain node $i$ \\
			\hline
			$\mc{L}$ & number of preferentially attached nodes \\
			\hline $\mc{L}_2$ & number of triad formations in $\mc{L}$\\
			\hline
			$W$ & weighted adjacency matrix of network\\
			\hline $\alpha$ &  gradient-tracking rate \\
			\hline			
			$W_{ij}$ & link weight beween nodes $i$ and $j$ \\
			\hline $\nabla f_i(\cdot)$ &  gradient of function $f_i(\cdot)$ \\
			\hline			$\overline{W}$ & Laplacian matrix \\
			\hline $D$ &  diagonal degree matrix \\
			\hline			
           $\mb{x}_i$ & optimization state variable at node $i$ \\
           \hline $\mb{y}_i$ &  auxiliary state variable at node $i$ \\
           \hline			
		\end{tabular}
		\label{tab_notation}
	\end{table}

\section{The Graph Theory Background} \label{sec_graph}
In this work, we use existing synthetic random networks to model real-world complex systems and tune the network clustering manually. This section introduces these random graphs and the metric to quantify network clustering. 
\subsection{Random Network Models}
Random SF networks are used to model large-scale complex systems because they capture important characteristics observed in many real-world networked systems. The main reason is the so-called SF property. In many real-world networks, such as social networks, IoT networks, transportation networks, and the World Wide Web, the degree distribution follows a scale-free pattern. This means that a few nodes (hubs) have a significantly higher degree than the majority of nodes. Random SF networks replicate this property, where the degree distribution follows a power-law distribution. This is crucial for capturing the heterogeneity and robustness observed in many complex systems. Moreover, random SF networks often exhibit a small-world topology, meaning that the average path length between any two nodes is relatively short. 

The most well-known model for random SF networks is the Barabási-Albert (BA) graph model \cite{barabasi1999emergence}. It is a generative model that explains how networks with SF properties can evolve through a process known as \textit{preferential attachment}. Starting with a small seed graph, the network grows by adding new nodes connecting to the old nodes. The preferential attachment, with $\mc{L}$ as the number of new links, means that new nodes are more likely to connect to existing nodes that already have a higher degree (more connections). This mechanism mimics the real-world scenario where popular nodes (hubs) attract more connections, e.g., in social networks. It is known that the BA degree distribution evolves to follow a power-law distribution.

\begin{figure} 
\centering
\includegraphics[width=2.2in]{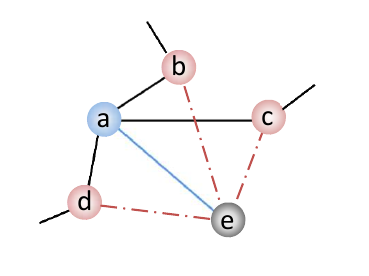}
\caption{This figure illustrates the triad formation in HK model to increase the network clustering. The newly added node 'e' makes connection to the preferentially attached node 'a' and connections to the $\mc{L}_2$ neighbours of 'a' (in this example $\mc{L}_2=3$) to make triangles (or triads). This directly increases the clustering. }  \label{fig_triad} 
\end{figure}

\begin{figure*} 
\centering
\includegraphics[width=3.2in]{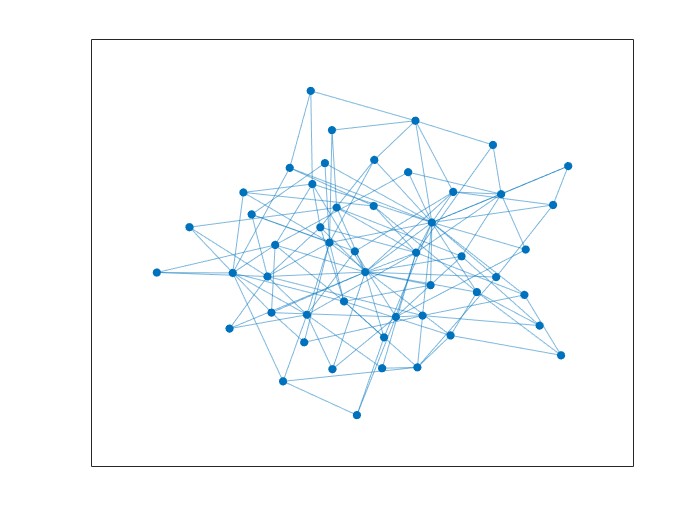}
\includegraphics[width=3.2in]{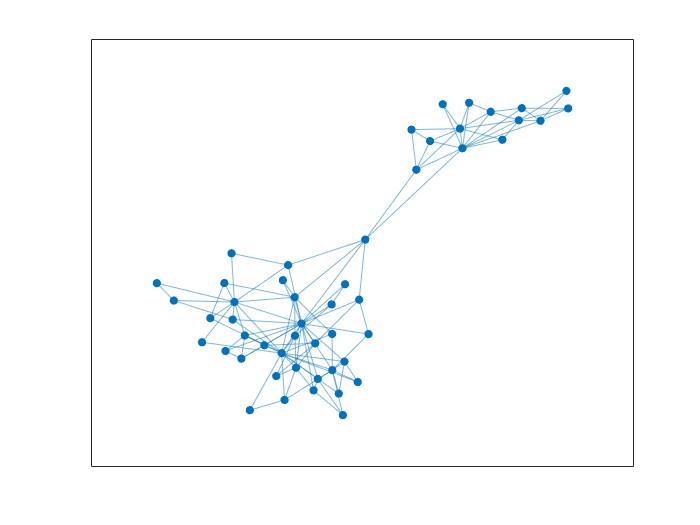}  		
\caption{This figure shows examples of SF network via BA model (Top) and CSF network via HK model (Bottom). Both network topologies have the same number of links and average node degree. The CSF network contains more triangles (or triads) as compared to the SF network. }  \label{fig_Sf} 
\end{figure*}

The Holme-Kim (HK) model is an extension of the (BA) model that introduces clustering into the evolving network structure \cite{holme2002growing} to address the observed lack of clustering in traditional SF networks generated by the BA model. The procedure is similar to the BA algorithm except for introducing a new feature called \textit{triadic closure} or \textit{triad formation}, which promotes the formation of triangles (clustering) in the network. After a new node connects to an existing node, there is an additional probability that the new node will also connect to a neighbour of the existing node, creating a triangle. This is better illustrated in Fig.~\ref{fig_triad}. 
The number of connected neighbours of the preferentially attached node is denoted by $\mc{L}_2 < \mc{L}$. This mechanism is inspired by the observation that many real-world networks exhibit a higher level of clustering than what is typically produced by the BA model \cite{klemm2002highly}. The generated CSF network by HK model exhibits the same power-law degree distribution, average node degree, and logarithmically
increasing average shortest-path length. Two sample networks generated by the BA and HK model are compared in Fig.\ref{fig_Sf}.	

\subsection{Network Clustering Coefficient}
The network clustering coefficient (NCC) is a measure used in network science to quantify the degree to which nodes in a network tend to cluster together. It provides insights into the local connectivity or the level of triadic closure in a network. In simple terms, it measures the likelihood that two neighbours of a node are also connected to each other. A high clustering coefficient indicates a high level of local cohesion or clustering within the network. The local NCC at node $i$ is defined as follows \cite{newman2003structure}:
\begin{align}
\mc{C}_i = \frac{2 \mc{T}_i}{d_i(d_i - 1)},
\end{align}
where $\mc{T}_i$ denotes the number of triads that includes node $i$ and $d_i$ is the degree of node $i$. The global NCC is then defined as
\begin{align}
\mc{C} = \frac{1}{n} \sum_{i=1}^n \mc{C}_i,
\end{align}		
with $n$ as the number of nodes (the network size). These formulas provide a way to quantify how tightly connected the neighbours of a node are, and the average of these values gives an overall measure of how much the network exhibits clustering.	
Note that the NCC is a normalized measure, ranging from $0$ to $1$. A large coefficient $\mc{C} \rightarrow 1$ indicates a high level of clustering, while a small coefficient $\mc{C} \rightarrow 0$ indicates a lack of clustering. There are approximations for the NCC of SF and CSF networks generated by the BA and HK models. We have \cite{szabo2003structural},
\begin{align} \label{eq_cba}
\mc{C}_{BA} \approx \frac{\mc{L}-1}{8} \frac{\log(n)^2}{n},
\end{align}		
with $\mc{L}$ as the number of links via preferential attachment. In other words, in the procedure of growing the seed network, each newly added node makes $\mc{L}$ connection to preferrentially attached nodes in the network. Further,
\begin{align} \label{eq_chk}
\mc{C}_{HK} \approx \frac{2\mc{L}_2}{d} + \frac{\mc{L}-1}{8} \frac{\log(n)^2}{n},
\end{align}	
with $d$ as the average node degree and $\mc{L}_2<\frac{d}{2}$ as the number of links to the neighbours of the preferentially attached node (triad formation). Both Eqs.~\eqref{eq_cba}-\eqref{eq_chk} hold for large values of $n \gg 1$ and $\mc{L} \gg 1$. Note that, for small networks, the clustering is closely tied with the choice of seed network and the mentioned formulas may not hold.

\section{Convergence-Rate vs. Clustering} \label{sec_conv}
\subsection{The Decentralized Optimization Algorithm}
We use the distributed learning algorithm proposed in  \cite{dsvm} (continuous-time) and  \cite{ddsvm} (discrete-time). The continuous-time dynamics is as follows:
\begin{align} \label{eq_xdot_g}	
\dot{\mb{x}}_i &= -\sum_{j=1}^{n} W_{ij} (\mb{x}_i-\mb{x}_j)-\alpha \mb{y}_i, \\ \label{eq_ydot_g}
\dot{\mb{y}}_i &= -\sum_{j=1}^{n} W_{ij} (\mb{y}_i-\mb{y}_j) + \frac{d}{dt} \nabla f_i(\mb{x}_i),
\end{align}
with $\alpha$ as gradient-tracking step-rate, $\mb{x}_i,\mb{y}_i \in \mathbb{R}^p$ respectively as the state and auxiliary variable at node $i$, and $W_{ij}$ as the weight of the link from node $j$ to node $i$. This networked optimization scenario is summarized in Algorithm~\ref{alg_1}.

\begin{algorithm} \label{alg_1}
\textbf{Given:}  cost function $f_{i}(\mb{x}_i)$, network topology $\mc{G}$, symmetric adjacency matrix  $W=[W_{ij}]$, gradient-tracking rate $\alpha$  \\	
\textbf{Initialization:} ${\mb{y}}_i(0)=\mb{0}_{p}$, random ${\mb{x}}_i(0)$
\\
\While{termination criteria NOT true}{
	Node $i$ finds local gradient $\boldsymbol{ \nabla} f_i(\mb{x}_i)$ \;
	Node $i$  receives variables $\mb{x}_j$ and $\mb{y}_j$ over network $\mc{G}$ \;			
	Node $i$ calculates Eqs.~\eqref{eq_xdot_g}-\eqref{eq_ydot_g} \;
}
\textbf{Return:}  optimal cost $F^*$\;	
\caption{Distributed learning at node $i$. }
\end{algorithm}

\subsection{Theoretical Convergence Analysis}
Let define the convergence Lyapunov function as ${V(\delta) = \frac{1}{2} \delta^\top \delta =  \frac{1}{2}\lVert \delta \rVert_2^2}$ with the variable $\delta$ denoting the difference of the state variable and the optimal solution, i.e.,
\begin{align}\delta  = \left(\begin{array}{c} {\mb{x}} \\ {\mb{y}} \end{array} \right) - \left(\begin{array}{c} {\mb{x}}^* \\ \mb{0}_{np} \end{array} \right) \in \mathbb{R}^{2np},
\end{align}
and $\mb{x}^*$ as the optimal state. This Lyapunov function represents the optimality gap of the proposed dynamics over time.
As proved in \cite{dsvm}, we have 
\begin{align} \label{eq_Re2}
\dot{V} = \delta^\top M(\alpha) \delta \leq \max_{1\leq j\leq p}\operatorname{Re}\{{\lambda}_{2,j}(\alpha)\} \delta^\top  \delta, 
\end{align}
with $\lambda_{2,j}(\alpha)$ as the largest nonzero eigenvalue of the following system matrix,
\begin{eqnarray} \label{eq_M}
M(\alpha) = \left(\begin{array}{cc} \overline{W} \otimes I_p & -\alpha I_{pn} \\ H(\overline{W}\otimes I_p) & \overline{W} \otimes I_p - \alpha H
\end{array} \right).
\end{eqnarray}
where $\overline{W}:= D-W$ (with $D$ as the degree matrix) is the Laplacian matrix containing the link weights and $H:=\mbox{diag}[\nabla^2 f_i(\mb{x}_i)]$ as the block-diagonal matrix of the second derivative of the local objective functions $f_i(\mb{x}_i)$. In fact, $\dot{V}$ defines the convergence rate of the optimality gap (the residual) along the time-evolution of the proposed dynamics \eqref{eq_xdot_g}-\eqref{eq_ydot_g}. 
Larger $|{\lambda}_{2,j}(\alpha)|$ implies faster decay of the optimality gap and higher convergence rate.
It is shown in \cite{dsvm} that ${\lambda}_{2,j}(\alpha)$ is directly related to the second largest eigenvalue of the Laplacian matrix $\overline{W}$. This value is referred to as the network \textit{algebraic connectivity}. On the other hand, as it is shown in \cite{alenazi2014cost}, the NCC and the algebraic connectivity are related. As a result, one can claim a relation between the convergence rate of Algorithm~\ref{alg_1} and the NCC. This is better illustrated by the simulation results in the next subsection. Note that analytical solutions to relate network structure and algebraic connectivity are proved to be NP-hard \cite{nagarajan2015maximizing}. Therefore, Monte-Carlo-based approaches are proposed to find the relation of the algebraic connectivity and different network properties in the literature \cite{alenazi2014cost,nagarajan2015maximizing}.

\subsection{Monte-Carlo Simulations for Convergence Analysis} \label{sec_sim}
\begin{figure} 
\centering
\includegraphics[width=3.5in]{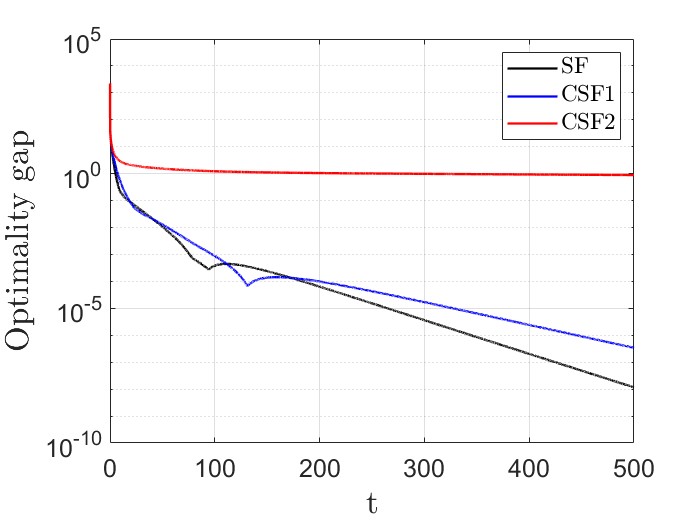}
\caption{This figure shows the decentralized learning with loss function given by Eq.~\eqref{eq_fij_sim} over different SF and CSF networks respectively modelled by BA and HK methods. It is clear that for the low-clustered SF network, the convergence is faster.}  \label{fig_sim_mc} 
\end{figure}

We consider the locally nonconvex function proposed by \cite{xin2021fast} for distributed machine learning. In this setup,  every node/machine $i$ optimizes the following:
\begin{align}\label{eq_fij_sim}
f_{i,j}(x_i) = 2 x_i^2 +3\sin^2(x_i)+a_{i,j} \cos(x_i) + b_{i,j}x_i,
\end{align}
where $\sum_{i=1}^n \sum_{j=1}^m a_{i,j} = 0$ and $\sum_{i=1}^n \sum_{j=1}^m b_{i,j}=0$ with nonzero parameters $a_{i,j},b_{i,j} \in (-1,1)$ (with $m=20$ data points). For Monte-Carlo simulation we consider randomly generated BA and HK networks of size $n=10000$ with three cases: (i) $\mc{L}=30$ (SF), (ii) $\mc{L}=30$, $\mc{L}_2=2$ (CSF1), and  (iii) $\mc{L}=30$, $\mc{L}_2=4$ (CSF2). The optimization parameters in Algorithm~\ref{alg_1} are set as $\alpha = 3$ and randomly weighted balanced adjacency matrix $W$. We average the cost evolution over $100$ Monte-Carlo network trials. The results are shown in Fig.~\ref{fig_sim_mc}. As it is clear, the convergence and learning over the SF network is considerably faster than its CSF counterparts.  
We compare the global clustering coefficient $\mc{C}$ for these three cases in Table~\ref{tab_1}.
\begin{table}[hbpt!]
\centering
\caption{ Comparing the network properties of the SF and CSF networks used in Fig.~\ref{fig_sim_mc}.}
\begin{tabular}{|l|c|c|c|c|}
	\hline
	Network&~ $\mc{L}$ & $\mc{L}_2$& $d$ & $\mc{C}$  \\
	\hline
	SF & ~$30$ & ~$0$ &~$57.1$  &~$0.031$   \\
	\hline
	CSF1 & ~$30$ & ~$2$ &~$57.2$ &~$0.102$  \\
	\hline
	CSF2 & ~$30$  & ~$4$ &~$56.2$ &~$0.172$  \\
	\hline
	\hline
\end{tabular}
\label{tab_1}
\end{table}
The data and results of Fig.~\ref{fig_sim_mc} and Table~\ref{tab_1} imply that for the low-clustered SF network the convergence is faster. Further, among the two other cases, the CSF1 network with lower clustering shows faster convergence as compared to the CSF2 network with higher clustering. As presented in Table~\ref{tab_1} the average node degree as another key network property is approximately the same for all three networks, implying the same amount of links among the nodes.

\begin{figure} 
\centering
\includegraphics[width=3.5in]{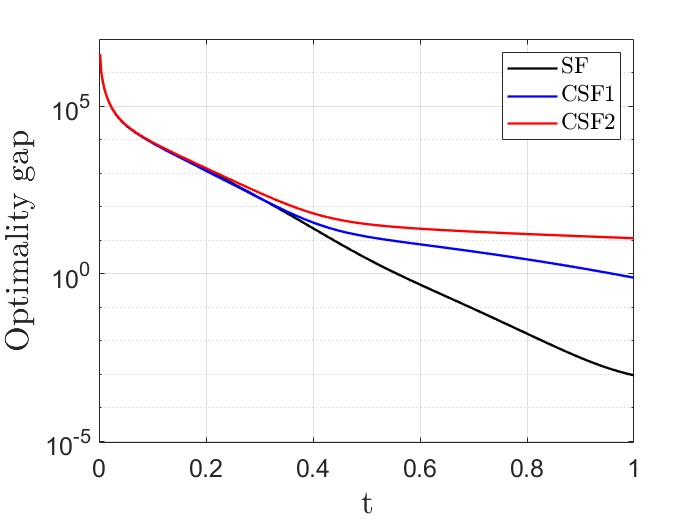}
\caption{This figure shows the decentralized optimization of the cost function given by Eq. \eqref{eq_fi_quad} over different network topologies. Clearly, the convergence over low-clustered networks is faster.}  \label{fig_sim_mc2} 
\end{figure}

Next, we consider a different convex objective function as proposed in \cite{doan2017distributed} for distributed resource allocation. The cost function is as follows: 
\begin{align}\label{eq_fi_quad}
f_{i}(x_i) = a_i(x_i-b_i)^4,
\end{align} 
with random nonzero parameters $a_i \in (0,0.025]$ and $b_i \in [-10,10]$. Monte-Carlo simulation (over $100$ trials) for BA and HK networks of size $n=10000$ are considered with three cases: (i) $\mc{L}=40$ (SF), (ii) $\mc{L}=40$, $\mc{L}_2=4$ (CSF1), and  (iii) $\mc{L}=40$, $\mc{L}_2=7$ (CSF2). We consider $\alpha = 1$ and random weights for the $W$ matrix. The optimality gap is averaged over $100$ Monte-Carlo trials and is shown in Fig.~\ref{fig_sim_mc2}.  
We compare the global clustering coefficient $\mc{C}$ for these three cases in Table~\ref{tab_2}. 
\begin{table}[hbpt!]
\centering
\caption{ Comparing the network properties of the SF and CSF networks used in Fig.~\ref{fig_sim_mc2}.}
\begin{tabular}{|l|c|c|c|c|}
	\hline
	Network&~ $\mc{L}$ & $\mc{L}_2$& $d$ & $\mc{C}$  \\
	\hline
	SF & ~$40$ & ~$0$ &~$76.6$  &~$0.041$   \\
	\hline
	CSF1 & ~$40$ & ~$4$ &~$76.1$ &~$0.145$  \\
	\hline
	CSF2 & ~$40$  & ~$7$ &~$75.1$ &~$0.223$  \\
	\hline
	\hline
\end{tabular}
\label{tab_2}
\end{table}   
Fig.~\ref{fig_sim_mc2} and Table~\ref{tab_2} again imply that convergence over low-clustered networks is faster than high-clustered ones.

\section{Case Studies: Real-World Networks } \label{sec_case}
In this section, we study the convergence rate and characteristics of some real-world networks given in \cite{konect,UCI,ASU}.
We compare the convergence rate and NCC for real-world networks as presented in Table~\ref{tab_case}. Recall from Eq.~\eqref{eq_Re2} that the convergence rate directly depends on $\operatorname{Re}\{{\lambda}_{2}(M(\alpha))\}$. The optimization cost is set the same as \eqref{eq_fi_quad} with similar parameters as in Section~\ref{sec_sim} and learning rate $\alpha = 0.001$.	
\begin{table*}[hbpt!]
\centering
\caption{ Comparing real-world networks: network characteristics versus clustering coefficient.}
\begin{tabular}{|l|c|c|c|c|c|}
	\hline
	Network Name&~ $n$ & $d$ & $\mc{C}$ & $\lambda_2(\overline{W})$ & $-\operatorname{Re}\{{\lambda}_{2}(M(\alpha))\}$ \\
	\hline
	Route-View & $6474$ & $4.088$ &$0.0059$  &$0.088$ & $0.084$  \\
	\hline
	Blogs & $10312$ & $64.77$ &$0.0324$ &$0.763$ & $0.75$ \\
	\hline
	Facebook(small) & $247$  & $7.61$ &$0.0489$ &$0.207$ & $0.197$  \\
	\hline
	US-Power-grid & $4941$  & $2.67$ &$0.103$ &$0.0008$ & $0.0012$ \\
	\hline
	Food-Web & $127$  & $16.69$ &$0.0057$ &$0.669$ & $0.91$ \\
	\hline
	Hamsterster-LLC & $1788$  & $13.95$ &$0.231$ &$0.103$ &  $0.032$\\
	\hline 
	Open-Flight & $2939$  & $10.38$ &$0.254$ &$0.0783$ & $0.0364$ \\
	\hline
	Elegans & $453$  & $8.973$ &$0.124$ &$0.264$ & $ 0.241$\\
	\hline
	Protein-Figeye & $2239$  & $2.88$ &$0.0353$  &~$0.0698$  & $0.0499$ \\
	\hline			
	Protein-Vidals &$2239$  & $4.10$ &$0.0076$  &$0.102$ & $0.0665$ \\
	\hline           
	Reactome & $6327$  & $23.32$ &$0.606$ &$0.0134$ &  $0.0123$\\
	\hline                            			
	\hline
\end{tabular}
\label{tab_case}
\end{table*}

\begin{figure} 
\centering
\includegraphics[width=3.3in]{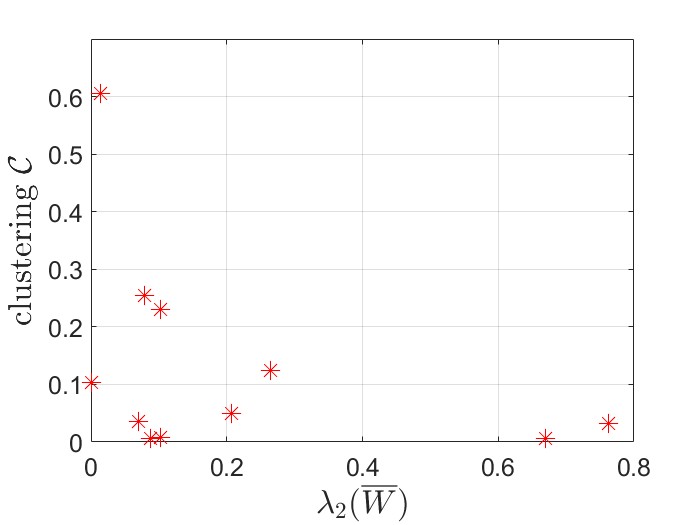}
\caption{This figure maps the clustering coefficient versus algebraic connectivity $\lambda_2(\overline{W})$ for the real networks in Table~\ref{tab_case}. It can be seen that, in general, larger algebraic connectivity is associated with lower clustering and vice versa.}  \label{fig_C_eigW} 
\end{figure}

\begin{figure} 
\centering
\includegraphics[width=3.3in]{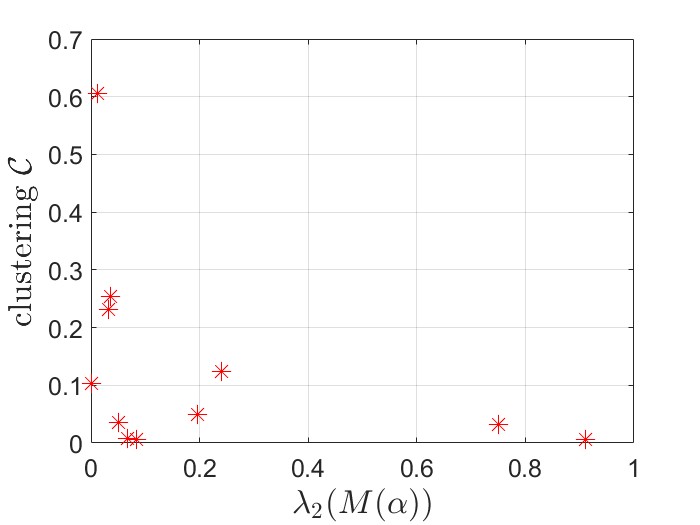}
\caption{This figure maps the clustering coefficient versus  $\lambda_2(M(\alpha))$ for networks in Table~\ref{tab_case}. $\lambda_2(M(\alpha))$ is a measure of the optimization convergence rate over the network.}  \label{fig_C_eigM} 
\end{figure}

To better illustrate, the results in table~\ref{tab_case} are shown in Fig.~\ref{fig_C_eigW} and Fig.~\ref{fig_C_eigM} to map the clustering coefficient $\mc{C}$ versus the algebraic connectivity $\lambda_2(\overline{W})$ and the eigenvalue $\lambda_2(M(\alpha))$. Recall that the convergence rate over the network can be quantified by $\lambda_2(M(\alpha))$. As it can be seen from Fig.~\ref{fig_C_eigM}, the network data implies that higher NCC is associated with low values of $\lambda_2(M(\alpha))$ and vice versa. It should be noted that other network characteristics (such as the number of links or average degree) may also affect the convergence rate. However, in general, one can see a trend in Fig.~\ref{fig_C_eigM} showing that highly-clustered networks are typically associated with lower convergence rates.


\begin{figure} 
	\centering
	\includegraphics[width=3.5in]{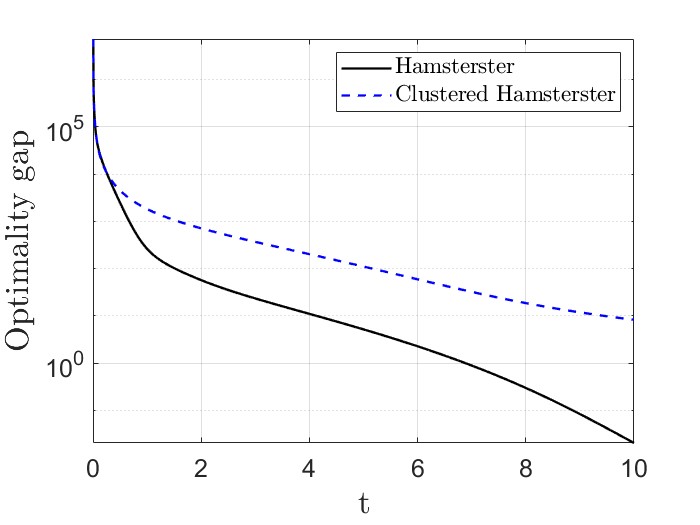}
	\caption{This figure compares the convergence rate of decentralized optimization over the Hamsterster-LLC network and its clustered version. The convergence over the high-clustered version is slower.}  \label{fig_sim_realnet} 
\end{figure}

Next, we consider the ``Hamsterster-LLC network'' as a real-world network \cite{konect} and tune (increase) its clustering by rewiring the links. As an example, one can use the rewiring strategy in \cite{liu2018optimization} to increase the clustering while keeping the degree distribution unchanged. Then, we compare the convergence rate of the optimization over the real network and its clustered version. The properties of the two networks are compared in Table~\ref{tab_3}.
	\begin{table}[hbpt!]
		\centering
			\caption{ Comparing the network properties of the Hamsterster-LLC network and its clustered version by rewiring the links.}
			\begin{tabular}{|l|c|c|c|}
				\hline
				Network&~ $n$ & $d$ & $\mc{C}$  \\
				\hline
				Hamsterster-LLC & $1788$  & $13.95$ &$0.231$    \\
				\hline
				Clustered Hamsterster-LLC & $1788$  & $14.06$ &$0.401$  \\
				\hline
				\hline
			\end{tabular}
			\label{tab_3}
	\end{table}   
	We apply the distributed optimization method by Algorithm~\ref{alg_1} to minimize the cost function~\eqref{eq_fi_quad} over both network topologies. The results are presented in Fig.~\ref{fig_sim_realnet}. As expected,  by increasing the clustering the convergence rate decreases.

\section{Concluding Remarks and Future Directions} \label{sec_conc}
Distributed and parallel data processing over networks has motivated this work to investigate the convergence properties and network characteristics.
Particularly, decentralized optimization strategies, as the main approach for recent parallelized ML approaches, are studied in this work. We investigate the relation between the convergence rate of these strategies and the clustering coefficient of the underlying network topology. This is done over both synthetic SF and CSF networks and real-world case studies.
Our results show that one can improve the convergence rate of decentralized optimization over networks by changing (reducing) their clustering coefficient. The tuning of the clustering coefficient can be performed by existing algorithms, for example via the results in \cite{kashyap2018mechanisms,fan2020hyperparameter}. This work opens many avenues of future research direction towards distributed ML, for example, to improve the convergence of decentralized optimization for error back-propagation in neural network training \cite{nedic2021distributed} or decentralized binary classification \cite{ddsvm}. Another future research direction is to improve the convergence rate of distributed resource allocation (or constraint-coupled optimization) over networked systems \cite{doan2017distributed,jiang2022distributed,scl}.
	
	\section*{Acknowledgements}
	 The authors acknowledge the use of some MATLAB codes from Koblenz Network Collection (KONECT) \cite{konect}.

	\bibliographystyle{spmpsci} 
	\bibliography{bibliography}
	
\end{document}